\documentclass{ifacconf}

\usepackage{graphicx}      
\usepackage{natbib}        
\usepackage{multirow}
\usepackage{comment}
\usepackage{url}
\usepackage[dvipsnames]{xcolor}
\usepackage{placeins}
\usepackage{amsmath}
\usepackage{float}

\begin{document}
\begin{frontmatter}

\vspace{-20pt}
\makebox[0pt][c]{\raisebox{1pt}{%
  \parbox{\textwidth}{\centering
    \small © 2025. This work has been accepted to IFAC for publication under a Creative Commons Licence CC-BY-NC-ND and will be presented at the Modeling, Estimation and Control Conference (MECC 2025) in Pittsburgh, Pennsylvania, USA.}
}}

\vspace{-20pt}

\title{Hydrodynamic Modeling Improvements for Floating Offshore Wind Turbines with Validation Results}

\thanks[footnoteinfo]{This work was supported by the U.S. Department of Energy
Advanced Research Projects Agency for Energy (ARPA-E) under
the grant DE-AR0001187.}

\author[First]{Doyal Sarker} 
\author[First]{Md Sakif} 
\author[First]{Tri Ngo}
\author[First]{Tuhin Das}

\address[First]{Department of Mechanical and Aerospace Engineering, University of Central Florida, FL 32816, USA (e-mail: doyal.kumar.sarker@ucf.edu, mdrafidulhaque.sakif@ucf.edu, tri.ngo@ucf.edu, tuhin.das@ucf.edu)}

\begin{abstract}   
This study presents key enhancements in hydrodynamic modeling using the strip-based Morison’s equation approach to enable rapid simulations of \textcolor{black}{Floating Offshore Wind Turbines (FOWT).} The modeling framework employs the relative form of the Morison equation, incorporating nonlinear irregular wave kinematics, vertical wave stretching, and diffraction corrections based on MacCamy-Fuchs (MCF) theory for large-scale,  non-slender structures. Wave kinematics are iteratively applied at dynamically displaced structural nodes to accurately capture fluid-structure interaction. Additionally, a discretization scheme is introduced to improve hydrodynamic load distribution across large horizontal structures of floaters. These enhancements are validated against experimental data from the Floating Offshore Wind and Controls Advanced Laboratory (FOCAL), which conducted a 1:70 scale test of the IEA-Wind 15MW reference turbine on the VolturnUS-S platform. Results demonstrate that the incorporation of nonlinear wave kinematics significantly improves low-frequency response accuracy. Furthermore, the vertical wave stretching and MCF corrections lead to surge response predictions that closely align with experimental measurements, while the improved load discretization significantly enhances heave and pitch response fidelity in wave-dominant frequency ranges. 
\end{abstract}


\end{frontmatter}
\section{Introduction}
\vspace{-6pt}
FOWTs present significant potential by enabling access to stronger and more consistent wind resources in deep-ocean environments, compared to their onshore counterparts. However, due to their structural complexity and dynamic interaction with the marine environment, the design and optimization of FOWTs heavily depend on high-fidelity numerical simulations \citep{carmo2024validation}. One of the most critical and challenging aspects of these simulations is the accurate modeling of hydrodynamic loads acting on floating platforms, which is essential for ensuring their stability, performance, and reliability \citep{hoeg2023semi}.
Hydrodynamic modeling approaches commonly used in FOWT analysis include potential-flow theory, the Morison equation, or a hybrid of both \citep{robertson2014definition,carmo2024validation}. Potential-flow theory provides solutions for radiation and diffraction effects but neglects viscous drag forces. In contrast, the Morison equation—an empirical formulation—captures radiation-induced added mass, wave excitation forces, and nonlinear viscous drag \citep{ishihara2019prediction}. A key advantage of the Morison equation is its capability to compute distributed hydrodynamic loads along the entire submerged structure, rather than approximating them as lumped forces at discrete locations.


Hydrodynamic models often utilize the Airy wave theory, which assumes that incident wave amplitudes are much smaller than their wavelengths, thereby neglecting the nonlinear effects of the real ocean environment. To address this limitation, higher-order wave models, such as the 2nd-order nonlinear wave model, have been adopted and highlighted in several studies \citep{duarte2014effects, zhang2020second,  roald2013effect}. When using the Morison equation, incorporating second-order difference and summation frequency wave kinematics generates second-order wave loads \citep{agarwal2011incorporating}. These second-order difference frequency wave loads primarily influence the slow-drift motion and low-frequency response of FOWTs.
In traditional hydrodynamic modeling using the Morison equation, wave loads are typically evaluated up to the mean sea water level (SWL), often neglecting loads at nodes between the SWL and the instantaneous wave surface. However, recent studies have emphasized the importance of accurately capturing wave loads above the mean SWL, highlighting the significance of nonlinear wave loads due to wave run-up and wave crest elevations. Ignoring the contribution of instantaneous free-surface fluctuations can result in underestimated hydrodynamic loads, particularly in extreme wave scenarios and high sea states encountered by FOWTs \citep{wang2022recent, wang2022oc6, carmo2024validation}. 

The potential flow hydrodynamic model, despite its accuracy, is computationally intensive, particularly when evaluating numerous structural configurations such as semi-submersible platforms \citep{hoeg2023semi}. In contrast, the strip-based Morison equation provides a computationally efficient alternative; however, its conventional form not account for diffraction effects, which become significant in large-scale, non-slender structures \citep{robertson2014definition}. To address this, \cite{hoeg2023semi} introduced a semi-analytical hybrid MacCamy-Fuchs-Morison (MCF-Morison) model, which can replace the potential flow model during preliminary design stage. For large-diameter cylindrical structures, where diffraction effects are significant, diffraction corrections derived from MCF theory \citep{maccamy1954wave} can be effectively integrated with the Morison equation. Typically, floating platforms are modeled as rigid bodies to evaluate global loads and responses. However, when internal structural load effects require detailed investigation, strip theory offers an effective solution. Strip theory facilitates the discretization of large rigid structures into multiple components \citep{wang2023methodology}, enabling a more accurate representation of spatial variations in wave kinematics and associated structural responses.
 
The discussion above highlights the effectiveness of the Morison equation in analyzing hydrodynamic loads on FOWTs with computational efficiency. However, several modifications are needed to update its underlying assumptions for improved accuracy. Therefore, this study aims to enhance the hydrodynamic modeling capabilities within the Control-oriented, Reconfigurable, and Acausal Floating Turbine Simulator (CRAFTS) developed by the authors, with these improvements validated against experimental data. CRAFTS is a causality-free modeling tool developed in the Modelica language for offshore wind turbine analysis. 
Previous hydrodynamic modeling efforts, including verification and validation for various platforms such as the OC3 spar-buoy and VolturnUS-S semi-submersible, are detailed in \cite{hasan2023stabilization} and \cite{sarker2024causality}, respectively. Additionally, the integration of aerodynamic, hydrodynamic, and control modules in CRAFTS is discussed in \cite{sarker2024modeling}. Hydrodynamic modeling within CRAFTS was limited to first-order wave kinematics using the conventional Morison equation. To provide deeper insights into the dynamic behavior of FOWTs and to develop a robust simulation platform, several critical enhancements have been integrated. This paper presents the formulations behind these enhancements. Furthermore, the implemented improvements are validated using experimental data from tests conducted on the VolturnUS-S semi-submersible platform during the FOCAL campaign \citep{wang2023experimental,robertson2023focal}. The experimental data is publicly accessible at: \textcolor{blue}{{https://a2e.energy.gov/ds/focal/focal.campaign4}}.

\section{Numerical Modeling}
\vspace{-6pt}
This section describes both the structural and the hydrodynamics model of the VolturnUS-S semi submersible floating planform. The structural model is numerically set up in CRAFTS in full scale. The hydrodynamics model, likewise is developed in CRAFTS, however, specific details are focused on formulation of modeling improvement.
\begin{figure}[h!]
\begin{center}
\includegraphics[trim= 0 0 0 0,clip, width=0.75\columnwidth]{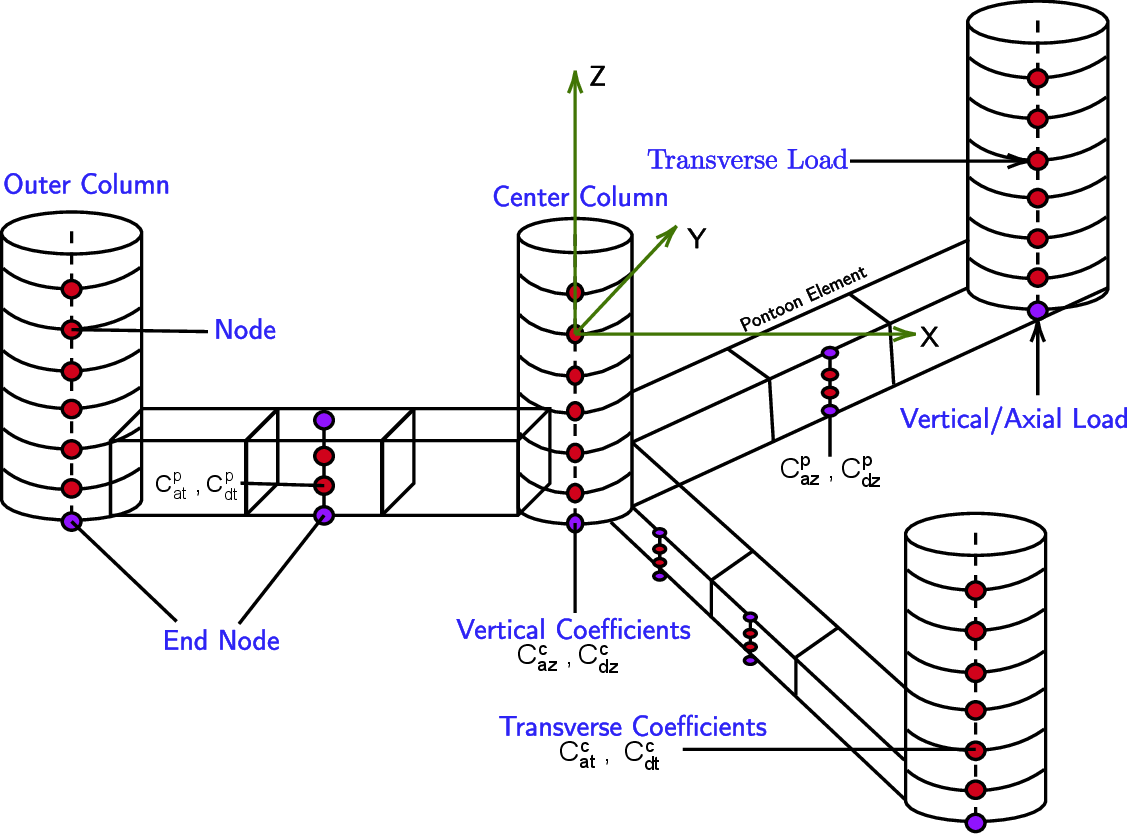}
\vspace{-8pt}
\caption{Coordinate system of Floating platform and discretization used in Morison equation} 
\label{fig:volturnus}
\end{center}
\end{figure}
\subsection{Structural Modeling}
\vspace{-5pt}
The VolturnUS-S semi-submersible platform comprises three vertical cylindrical outer columns connected to a central column by three rectangular pontoons, as shown in Fig.~\ref{fig:volturnus}. All structural components are modeled as rigid bodies, with each pontoon divided into three equal segments. As this study focuses on hydrodynamic modeling, aerodynamic effects are excluded. To accurately capture overall mass and inertia, the tower is modeled as a rigid body attached to the central column, and the rotor-nacelle assembly (RNA) is simplified as a point mass atop the tower, though these components are omitted from the figure for clarity. The platform is connected by three mooring lines, extending from the outer columns to fixed anchor points. These lines are modeled as massless linear springs, differing from conventional catenary modeling~\citep{noboni2025modeling}, and are assumed to have no hydrodynamic loading, resembling the fishing lines used in experiments. Additional mass or stiffness from sensor wiring bundles (umbilical) is not included, as experimental data with and without umbilical are available. Detailed geometry, inertia properties, and mooring characteristics are provided in \cite{robertson2023focal}.

\subsection{Hydrodynamics Modeling}
\vspace{-6pt}
The hydrodynamics module developed in the CRAFTS simulator uses the relative form of the Morison equation that accounts for radiation-induced added mass, wave excitation, and non-linear drag loads. In addition to those hydrodynamic loads, buoyancy and gravity loads are included as part of the linear hydrostatic loads. More details on the hydrodynamic modeling of the FOWT platform in CRAFTS simulator can be found in our previous work \citep{sarker2024causality}. In the present study, we introduce several key modeling practices and advanced formulations
to improve the accuracy, computational efficiency, and overall fidelity of the hydrodynamic model-extending its capabilities beyond those presented in earlier publications.

\subsubsection{Second-order wave and spatial effects}
In the CRAFTS hydrodynamics module, random waves were initially modeled as linear irregular waves using Airy wave theory, with associated wave kinematics computed up to mean sea water level. In this study, in addition to first-order wave kinematics, second-order kinematics—specifically difference-frequency and summation-frequency components—are incorporated in finite water depth following the theoretical formulation by \citep{agarwal2011incorporating}. For brevity, we only show the formulation of $2^{nd}$ order component, $\eta_2(x,t)$ of irregular wave elevation, $\eta(x,t) = \eta_1(x,t) + \eta_2(x,t)$. 
{\small
\begin{equation} \label{eqn:eta2}
\eta_2(x,t)=\text{Re}\left\{\sum_{m=1}^{N} \sum_{n=1}^{N} A_m A_n B_{mn}^{\pm} e^{i[(\omega_m \pm \omega_n)t-(k_m \pm k_n)x]}\right\}
\end{equation}
}
where $B^{\pm}_{mn}$ is the transfer function obtained from the solution of Laplace's equation. $A_m$ is the complex amplitude of $m^{th}$ wave and given by $A_m = a_m e^{i\phi_m}$. Here, $a_m$, $\omega_m$, $k_m$, and $\phi_m$ are the $m^{th}$ wave amplitude, frequency, wave-number, and random phase, respectively, can be directly obtained from wave spectrum.  Eqn.~\ref{eqn:eta2} represents second-order wave-wave interaction effects, capturing both sum-frequency $(\omega_m + \omega_n)$ 
and difference-frequency  $(\omega_m + \omega_n)$ wave components. The double summation in Eqn.~\ref{eqn:eta2} is computationally expensive, even for a moderate number, $N$. Therefore, this equation can be expressed in single summation and performed Inverse Fast Fourier Transform (IFFT) to obtain time-domain wave elevation as described in \cite{agarwal2011incorporating}. 
\begin{figure}[h!]
\begin{center}
\includegraphics[trim= 0 0 0 0,clip, width=1.0\columnwidth]{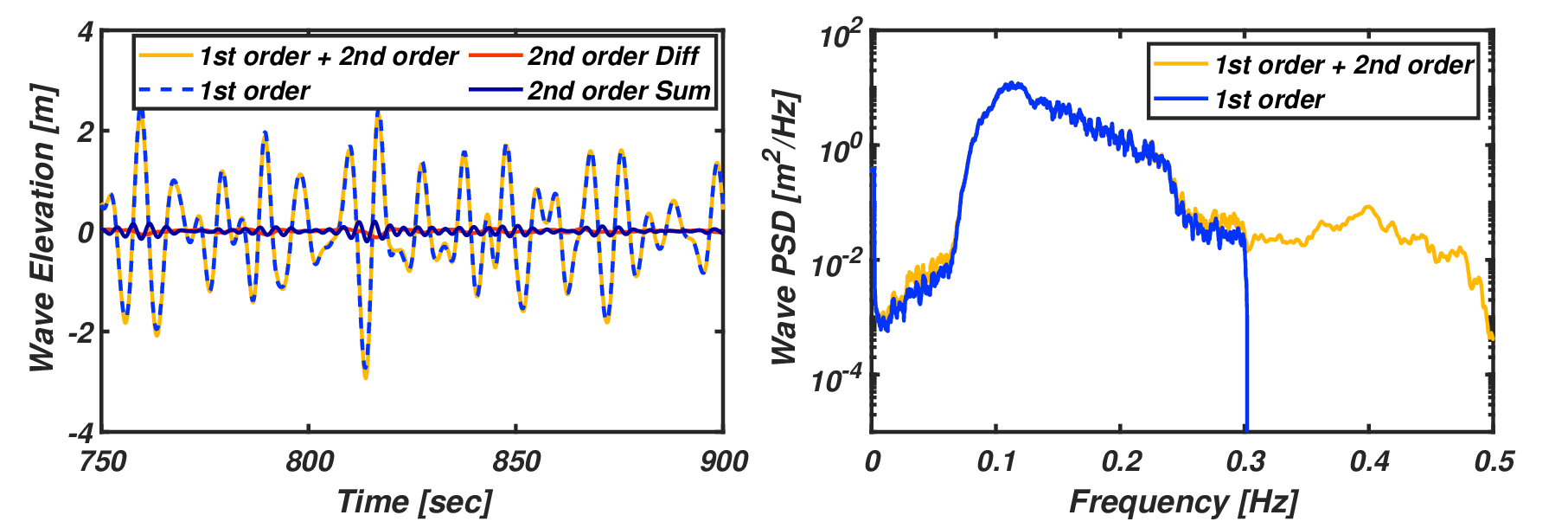}
\vspace{-15pt}
\caption{1st and 2nd order contribution in wave elevation shown in both time and frequency domains} 
\label{fig:2ndwave}
\end{center}
\end{figure}
To demonstrate the effect of nonlinear waves, we simulated a JONSWAP spectra irregular wave with a significant wave height of $H_s = 3.1 ~\text{m}$ and a peak period of $T_p = 8.96 ~\text{s}$ with a cut-off frequency of $0.3~Hz$. \textcolor{black}{Fig.~\ref{fig:2ndwave} shows the wave elevation over a 750–900 s time window for visualization purposes, and the Power Spectral Density (PSD) computed over the 500–9000 s interval, excluding the initial transient period.} While nonlinear effects are subtle in the time series, they are evident in the PSD, especially at low and high frequencies. The $1^{st}$-order wave PSD drops at the cut-off (~0.3 Hz) frequency, while the $2^{nd}$-order wave captures extra energy beyond the cut-off, due to sum-frequency effects, and enhances the low-frequency range via difference-frequency contributions. 

\begin{figure}[h!]
\begin{center}
\includegraphics[trim= 0 0 0 0,clip, width=0.70\columnwidth]{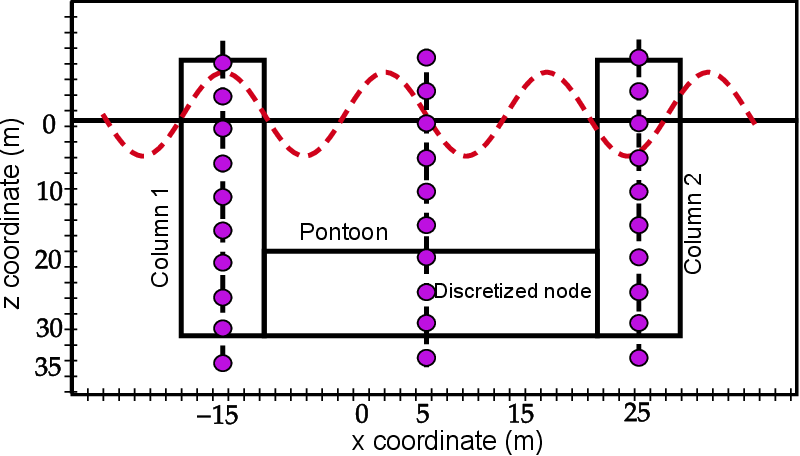}
\vspace{-5pt}
\caption{Representation of nodal points for calculating wave kinematics} 
\label{fig:spatial}
\end{center}
\end{figure}

To ensure the accurate application of wave kinematics in  the Morison equation, precomputed wave kinematics are interpolated at dynamically varying vertical positions $(z)$ as the structure moves. For instance, consider a floating structure with two vertical columns connected by a pontoon, as shown in Fig.~\ref{fig:spatial}. For Column 1, wave kinematics are initially computed at an undisplaced horizontal nodal location $x$ (e.g., $x = -15)$ across a predefined vertical range $z$ (e.g., $z = -35~to~15$), with sufficient margin to account for motion. During simulation, these precomputed values are  interpolated at each time step to evaluate wave kinematics at updated $z$-positions, ensuring accurate hydrodynamic force calculations that reflect structural orientation and motion. The horizontal nodal location $(x)$ are then iteratively updated based on the hydrodynamic loads from the previous step, maintaining consistency between structural motions and wave kinematics.

\vspace{-5pt}
\subsubsection{Wave Stretching and MCF correction} \label{sec:MCF}
\begin{figure}[h!]
\begin{center}
\includegraphics[trim= 0 0 0 0,clip, width=0.70\columnwidth]{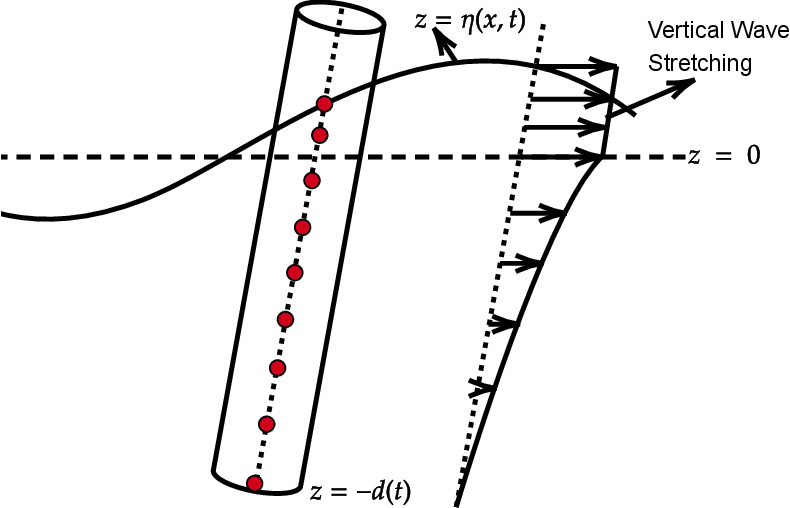}
\vspace{-5pt}
\caption{Conceptual sketch of vertical wave stretching} 
\label{fig:vws}
\end{center}
\end{figure}

In a simplified modeling approach, wave kinematics and the resulting wave loads are often evaluated up to the mean SWL. This assumption can lead to inaccuracies in predicting wave kinematics and in-line wave load, as it neglects the region between the SWL and the instantaneous wave surface, $\eta (x,t)$. To accurately capture wave kinematics in this region, wave stretching methods are employed. Common techniques include: (1) Wheeler stretching (2) Vertical stretching (3) Extrapolation stretching. As part of enhancing the hydrodynamic modeling capabilities, this study implements vertical wave stretching - a straightforward method that assumes the wave kinematics from the SWL up to $\eta (x,t)$ remain constant and equal to those at SWL $(z=0)$, as shown in Fig.~\ref{fig:vws}.

\vspace{-5pt}
In hydrodynamic modeling of large-diameter circular cylinders - commonly used in offshore platforms - diffraction effects become significant when the ratio of cylinder diameter $(D)$ to wavelength $(L)$ exceeds 0.2 \citep{faltinsen1993sea}. To address diffraction effects in such cases, this study adopts the MCF theory, which is particularly effective in accounting for diffraction in short-wavelength regimes. The conventional Morison equation, which applies a constant effective inertia coefficient, $C_M$ $(C_M = 1 + C_a$, and $C_a$ is the added mass coefficient), often overpredicts wave excitation loads \citep{wang2022recent}. MCF theory addresses this limitation by modifying the constant effective inertia coefficient as follows:
\begin{equation} \label{eqn:CM(ka)}
   \scalebox{0.90}{$ C_M (ka) =  \frac{4}{\pi (ka)^2\sqrt{A(ka)}}$}
\end{equation}
where $k$ is the wave-number and $a$ is the radius of circular cylinder. $A(ka)$ is given by:
\begin{equation} \label{eqn:A(ka)}
    \scalebox{0.90}{$ A(ka) = {J'_1}^2 (ka) + {Y'_1}^2 (ka) $}
\end{equation}
$J'_1$ and $Y'_1$ are the derivative of the Bessel function of the first kind of order 1 and derivative of the Bessel function of the second kind of order 1 respectively. The associated phase $(\alpha )$ is defined as a function of $ka$,  \textcolor{black}{$\alpha(ka) = tan^{-1}({J'_1} (ka) / {Y'_1} (ka))$}. This MCF correction is applied specially during the calculation of transverse wave acceleration  $(\Dot{u}_t)$, and only for surface piercing vertical cylinder or column. The modified wave acceleration $(\Dot{u}_t^{mod})$ is then calculated as:
\begin{equation}
    \Dot{u}_t^{mod}(x,z,time) = C_M(ka)\Dot{u}_t(x,z,time)
\end{equation}
It should be noted that the original equation for $\Dot{u}_t(x,z,time)$ remains unchanged. However, $\Dot{u}_t^{mod}(x,z,time)$ incorporates the phase $(\alpha(ka))$ introduced by the MCF theory.

\begin{figure}[h!]
\begin{center}
\includegraphics[trim= 0 0 0 0,clip, width=0.7\columnwidth]{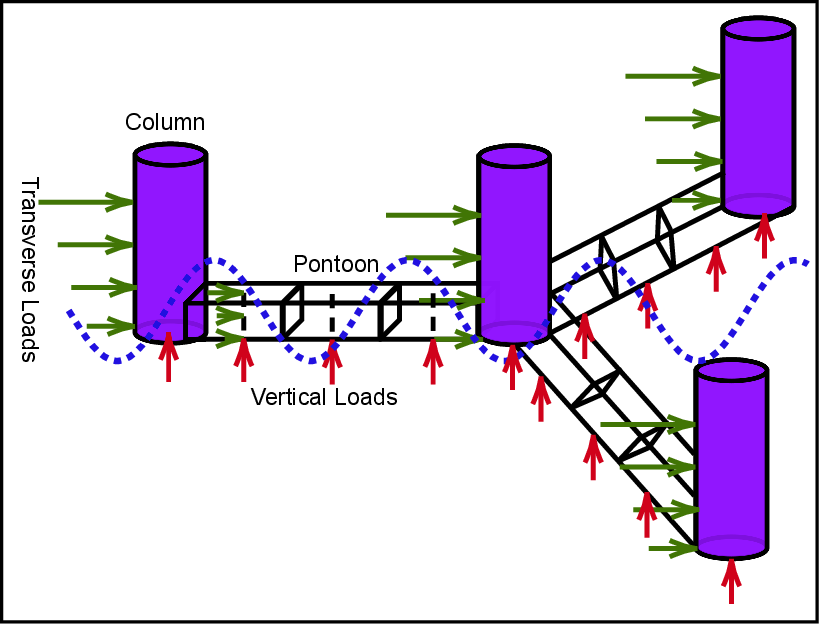}
\vspace{-5pt}
\caption{Illustration of pontoon discretization and strip-theory load distribution on different elements} 
\label{fig:pontoondis}
\end{center}
\end{figure}
\subsubsection{Pontoon Load Discretization}
We extend the strip theory approach to model the hydrodynamics for any pontoon structure typically used in semi-submersible floating platform e.g., VolturnUS-S, WindMoor Platforms. In our previous work \citep{sarker2024causality}, we modeled pontoon hydrodynamics as a single rigid body and lumped the loads at the single center point located at the bottom face of the pontoon. However, a single point calculation for a large pontoon may not capture wave-induced loads accurately, especially the vertical heave force, due to the spatial variation and phase difference of wave kinematics. In long-wavelength case, part of the large pontoon might be in the wave-crest region, while another part in the wave-trough region.  Therefore, to better study the loads distribution along the pontoon, we divided the rigid pontoon into multiple small rigid pontoon elements that accounts the spatial variation in wave kinematics and results different hydrodynamics loads at different pontoon elements. The application of the load distribution is illustrated in Fig.~\ref{fig:pontoondis}. In the current study, the pontoon of the VolturnUS-S semi-submersible platform is divided into three small pontoons with a length of $11.5~m$ of each element. It should be worth mentioning that there's no need to compute the resultant loads from individual pontoon load contributions, as CRAFTS simulator inherently accounts for this automatically. 
\vspace{-5pt}

\section{Results and discussion}
\vspace{-5pt}
This section validates the enhanced hydrodynamic modeling of the VolturnUS-S floating platform, as implemented in CRAFTS. The FOCAL experimental campaigns were conducted at the Alfond Wind-Wave Ocean Engineering Laboratory (W2) at the University of Maine, using a 1:70 Froude-scaled model of the FOWT. In both numerical simulations and experimental tests, the X-axis is aligned with the 0° wave heading, the Z-axis points vertically upward, and the Y-axis is defined according to the right-hand rule. In this study, irregular wave propagation is considered at a 0° heading; therefore, only the significant dynamic responses in surge, heave, and pitch are discussed. Additionally, due to the platform's geometric symmetry, only surge, heave, and pitch motions are considered in the free decay test.
\vspace{-5pt}

\subsection{Hydrodynamic Coefficients Calibration}
\vspace{-5pt}
A series of free decay tests are first conducted to calibrate key hydrodynamic coefficients, including added mass ($C_a$), quadratic drag ($C_d$), and additional linear damping ($B$). These coefficients are calibrated through an automatic tuning process using a Genetic Algorithm (GA) optimization framework. \textcolor{black}{The objective function is defined as: $J(\boldsymbol{x}
) = \sum w_i \, Xi_{\text{RMSE}}$}. Where,
\[
\scalebox{0.80}{$
X_{\text{RMSE}} = \sqrt{\frac{1}{N} \sum_{k=1}^{N} \left(X_k^{\text{sim}} - X_k^{\text{exp}}\right)^2}
$}
\]

\textcolor{black}{$\boldsymbol{x}$
 is the design variable vector of hydrodynamic coefficients. Index \( i = 1,2,3 \) denotes surge, heave, and pitch, \( w \) is the weight, and \( X \) is the time series response. Root-Mean-Square-Error (RMSE) is computed between simulation \( (X^{\text{sim}}) \) and experimental \( (X^{\text{exp}}) \) free decay data in time domain.} While the details of the optimization procedure are beyond the scope of this study and will be presented in a separate publication. The optimization ensures satisfactory reproduction of the natural periods and an appropriate balance between linear (P) and quadratic (Q) damping contributions in calm water. Additional linear damping is introduced into the system to compensate for the absence of radiation damping in the current Morison equation. This additional damping improves agreement with experimental data, particularly for small-amplitude motions observed during free decay tests. The added mass and drag coefficients are  specific to each structural member and direction, therefore, distinct sets of coefficients are calibrated for each structural member and listed in Table \ref{tb:decaytuned}. These decay-tuned values are directly applied in the irregular wave simulations.
\begin{table}[hb]
\begin{center}
\caption{Free decay tuned coefficients}\label{tb:decaytuned}
\scalebox{0.85}{
\begin{tabular}{ccccccc}
\hline
Member & \multicolumn{3}{c}{Added mass} & \multicolumn{3}{c}{Quadratic drag} \\\hline
 & $C_{at}$ & & $C_{az}$ & $C_{dt}$ & & $C_{dz}$ \\\hline
Column & 1.50 &  & 0.12 & 2.09 & & 2.23 \\
Pontoon & 0.11 & & 1.87 & 1.84 & & 4.96\\\hline
& \multicolumn{6}{c}{Global linear damping, $B$} \\
& \multicolumn{2}{c}{Surge, $B_{11}$} & \multicolumn{2}{c}{Heave, $B_{33}$} & \multicolumn{2}{c}{Pitch, $B_{55}$} \\\hline
Global & \multicolumn{2}{c}{4.4E4 Ns/m} & \multicolumn{2}{c}{1.6E5 Ns/m} & \multicolumn{2}{c}{3.63E8 Nms/rad} \\\hline
\end{tabular}}
\end{center}
\end{table}
\vspace{-10pt}


\subsection{Irregular Wave Testing}
\vspace{-5pt}
This study aims to demonstrate the improved hydrodynamic modeling of the FOWT under different wave conditions by comparing numerical results with experimental data. In the improved model, the platform is subjected to second-order irregular wave excitation. The column members are modeled using the MCF diffraction-corrected approach with vertical wave stretching (VWS) enabled. Additionally, each pontoon is subdivided into three equal rigid elements, with hydrodynamic loads computed individually to account for the spatial variation of wave kinematics. The CRAFTS simulation cases are categorized as follows:

\begin{itemize} \item \textbf{Case 1:} $1^{st}$-order wave only
\item \textbf{Case 2:} $2^{nd}$-order wave 
\item \textbf{Case 3:} $2^{nd}$-order wave + VWS + MCF
\item \textbf{Case 4:} Case 3 + Pontoon load discretization
\end{itemize}

\begin{figure}[h!]
\begin{center}
\includegraphics[trim= 0 0 0 0,clip, width=0.95\columnwidth]{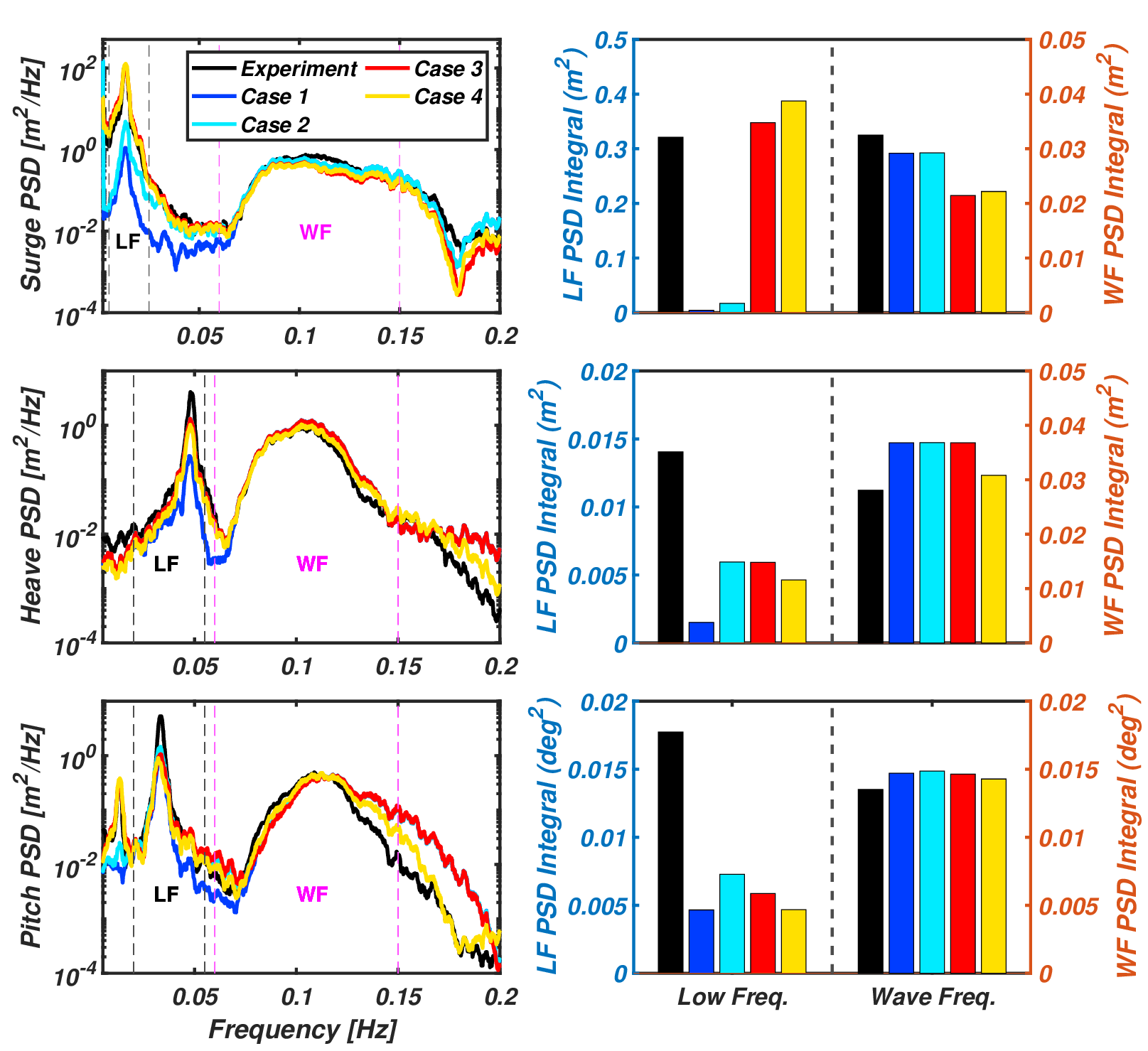}
\vspace{-5pt}
\caption{PSD (left) and PSD integral (right) for LF and WF range comparing with experiment in irregular wave 1} 
\label{fig:lc31}
\end{center}
\end{figure}

Fig.~\ref{fig:lc31} shows the power spectral density (PSD) of the surge, heave, and pitch responses of the VolturnUS-S floating platform under Irregular Wave 1 ($H_s = 3.1,\text{m}$, $T_p = 8.96,\text{s}$). The platform's dynamic responses are predominantly influenced by two distinct frequency regions: the wave-excitation frequency band, associated with first-order wave loading, and the resonance frequency band, resulting from nonlinear difference-frequency wave loading. For quantitative comparison, these regions are categorized as low-frequency (LF) and wave-frequency (WF) ranges. The LF range is defined as follows: surge—$0.005$ to $0.025,\text{Hz}$, heave—$0.02$ to $0.055,\text{Hz}$, and pitch—$0.02$ to $0.055,\text{Hz}$. The WF range is defined as $0.06$ to $0.15,\text{Hz}$ for each response. These LF and WF ranges are also highlighted by vertical dashed lines in Fig.~\ref{fig:lc31} (left) to aid visual interpretation.

\begin{table}[h]
\begin{center}
\caption{Summary of LF and WF PSD integrals error (Irregular Wave 1)}\label{tb:Error}
\scalebox{0.9}{
\begin{tabular}{c|ccc|ccc}
\hline
& \multicolumn{3}{c|}{Low-Frequency (LF)} & \multicolumn{3}{c}{Wave-Frequency (WF)} \\\hline
Case & Surge & Heave & Pitch & Surge & Heave & Pitch \\\hline
Case 1 & -99\% & -89\% & -74\% & -10\% & 31\% & 9\% \\\hline
Case 2 & -95\% & -58\% & -59\% & -10\% & 31\% & 10\% \\\hline
Case 3 & 8\% & -58\% & -67\% & -34\% & 31\% & 8.4\% \\\hline
Case 4 & 21\% & -67\% & -74\% & -32\% & 10\% & 6\% \\\hline
\end{tabular}}
\end{center}
\end{table}

The incorporation of key hydrodynamic modeling enhancements improves agreement with experimental data compared to the baseline modeling framework using the $1^{st}$-order wave model. While the $1^{st}$-order model performs reasonably well in the WF range, the inclusion of $2^{nd}$-order wave kinematics significantly improves the LF response—particularly in heave and pitch—by capturing nonlinear difference-frequency loads that excite resonance modes (Fig.~\ref{fig:lc31}). Pitch PSD analysis reveals that only the $1^{st}$-order model can not capture the peak near the surge natural frequency. Case 3, which includes VWS and MCF diffraction corrections, significantly improves the LF surge response. This improvement is also observed in the pitch response due to the strong coupling between surge and pitch motions. These enhancements primarily influence the column members in the transverse direction, which explains the limited impact on the heave response. Additionally, vertical hydrodynamic loads are applied only at structural ends (Fig.~\ref{fig:volturnus}). However, in Case 3, the WF surge response exhibits greater deviation from the experiment data. This discrepancy arises from the nature of the MCF correction. As discussed in Section~\ref{sec:MCF}, constant added mass models tend to overpredict wave-excitation loads, particularly at higher frequencies. The MCF model mitigates this by reducing the effective inertia coefficient as wave frequency increases (or wavelength decreases). Finally, incorporating pontoon load distribution to account for spatial wave effects further improves agreement with experimental data, particularly in the WF heave and pitch responses (see Fig.~\ref{fig:lc31}, PSD and PSD integral plots). \textcolor{black}{However, a fine load distribution of pontoon may affect the LF surge response.}
A summary of quantitative analysis on error based on the PSD integrals is shown in Table \ref{tb:Error}. 

\begin{figure}[h!]
\begin{center}
\includegraphics[trim= 0 0 0 0,clip, width=0.95\columnwidth]{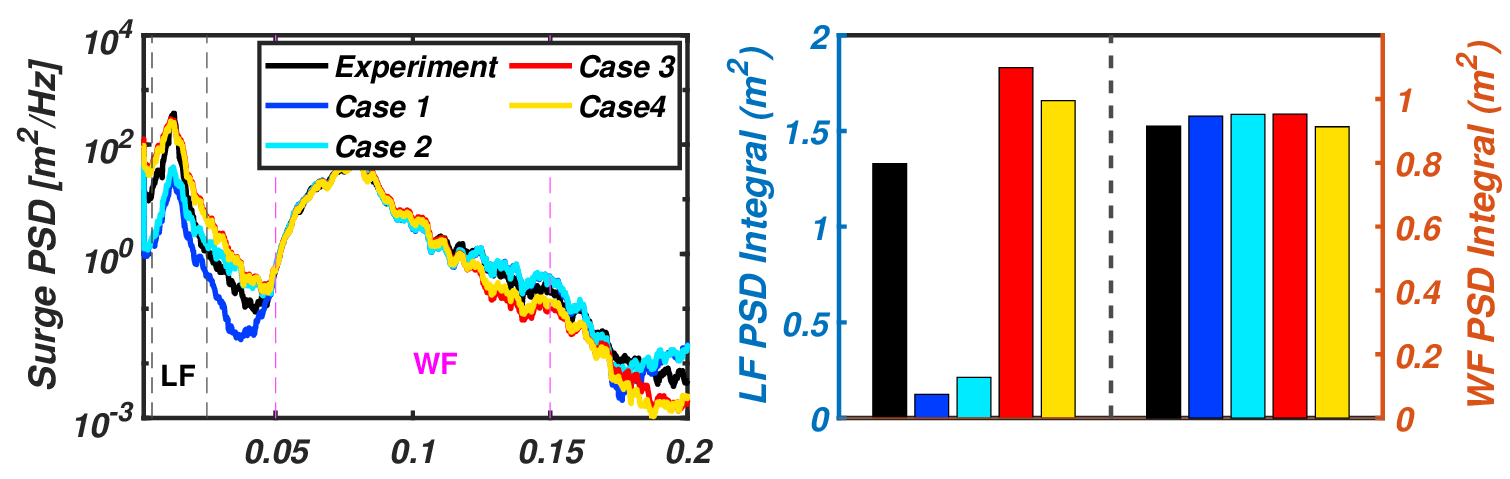}
\includegraphics[trim= 0 0 0 0,clip, width=0.95\columnwidth]{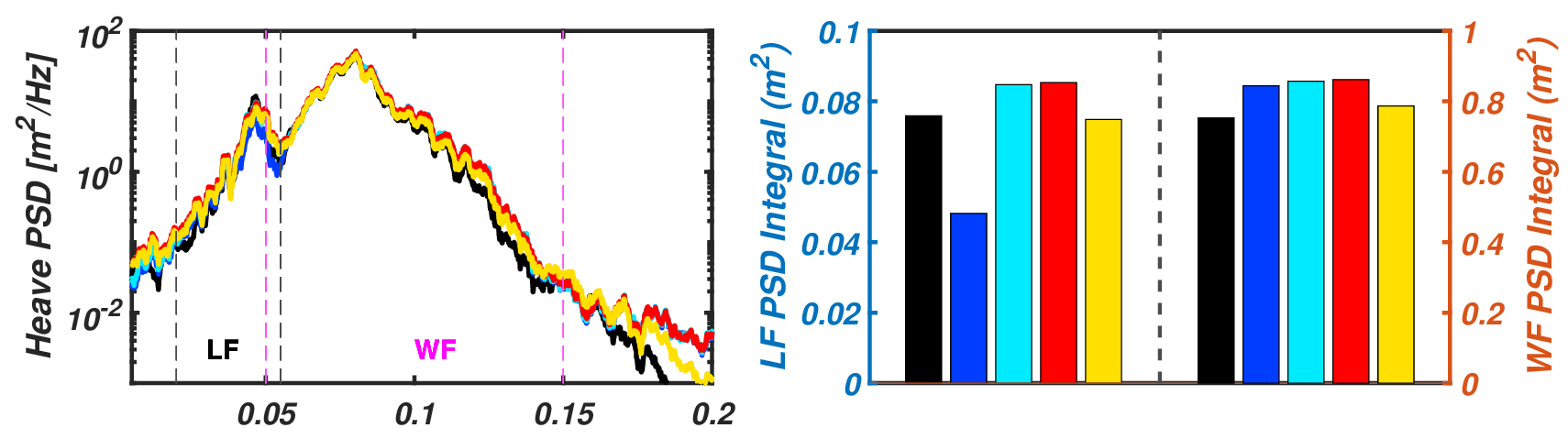}
\includegraphics[trim= 0 0 0 0,clip, width=0.95\columnwidth]{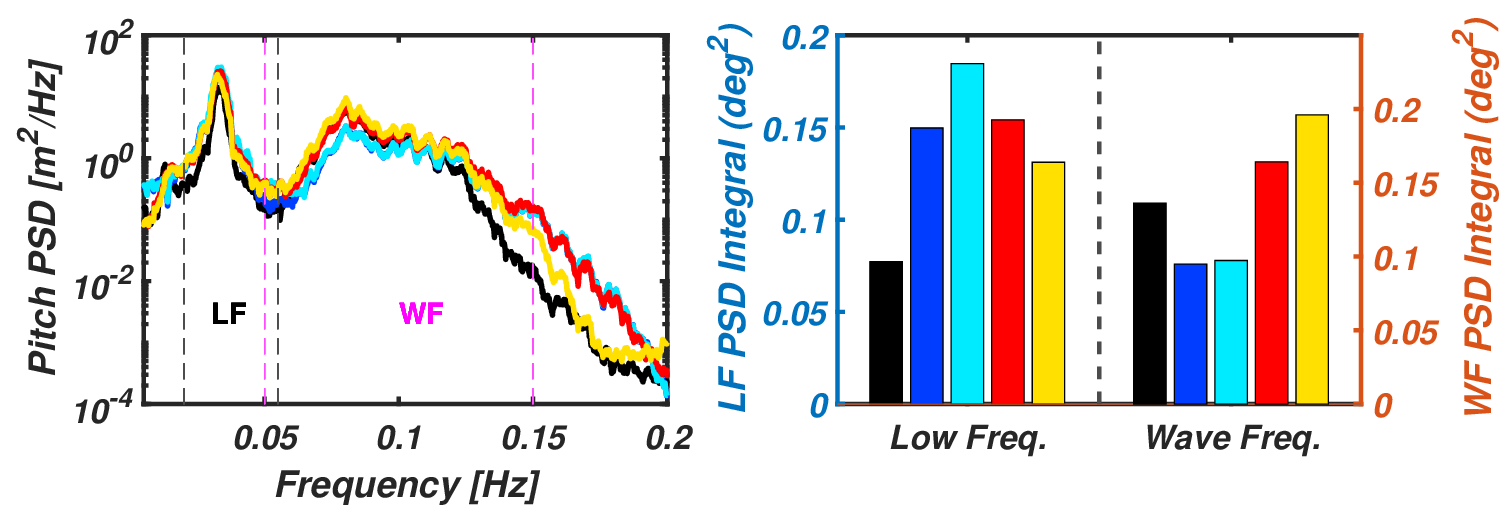}
\vspace{-5pt}
\caption{PSD (left) and PSD integral (right) for LF and WF range comparing with experiment in irregular wave 2} 
\label{fig:lc32}
\end{center}
\end{figure}
Another irregular wave case (Irregular Wave 2: $H_s = 8.1m$, $T_p = 12.8s$) representing a higher sea state is simulated and compared with experimental data in Fig.~\ref{fig:lc32}. While the LF range remains unchanged, the WF range is adjusted to 0.05–0.15 Hz due to the larger peak period. As with Irregular Wave 1, progressive modeling improvements from Case 1 to Case 4 are evident in Fig.~\ref{fig:lc32}, although these are not detailed here for brevity. Notably, the Morison equation provides improved WF surge predictions in this higher sea state. This improvement is due to the fact that, at longer wavelengths (associated with higher sea states), the Morison equation offers a better approximation. Analysis of the LF responses in Fig.~\ref{fig:lc31} and Fig.~\ref{fig:lc32}, along with the LF PSD integral errors shown in Table~\ref{tb:Error}, reveals a notable 
over- or under- prediction in surge, heave, and pitch responses. This discrepancy can be attributed to the direct use of decay-tuned damping coefficients in the irregular wave simulations. While these coefficients are calibrated using free decay tests, they may not fully capture the energy dissipation characteristics under irregular wave testing. Recent studies suggest that damping coefficients should be recalibrated based on specific wave conditions to improve prediction accuracy \citep{wang2025oc7}. Although this refinement is beyond the scope of the present work, the authors intend to incorporate sea state-dependent damping calibration in future studies using optimization-based approaches. 
\vspace{-8pt}

\section{Conclusion}
\vspace{-5pt}
This study demonstrates the effectiveness of several enhancements in hydrodynamic modeling of FOWTs using Morison's equation. By systematically incorporating second-order wave kinematics, vertical wave stretching, MacCamy–Fuchs diffraction correction, and spatially distributed pontoon loads, the model shows progressive improvement in capturing the FOWT’s dynamic responses in both the wave-frequency and low-frequency ranges. The inclusion of the second-order wave model significantly improves low-frequency predictions, particularly for heave and pitch motions, underscoring the importance of nonlinear difference-frequency wave loads. Vertical wave stretching further enhances wave kinematics by accounting for wave loads up to the instantaneous wave surface, improving surge response accuracy. Additionally, the MCF diffraction correction adjusts the added mass as a function of wavelength, mitigating the overprediction of wave-excitation loads at high frequencies through wavelength-dependent diffraction effects. The use of damping coefficients tuned from free decay tests introduces discrepancies in irregular wave testing, particularly at low frequencies, highlighting the need for sea-state-dependent damping calibration. Future work will address this issue using optimization-based methods to further refine the model’s fidelity.

\bibliography{ifacconfref}           

\begin{thebibliography}{20}
\providecommand{\natexlab}[1]{#1}
\providecommand{\url}[1]{\texttt{#1}}
\providecommand{\urlprefix}{URL }
\expandafter\ifx\csname urlstyle\endcsname\relax
  \providecommand{\doi}[1]{doi:\discretionary{}{}{}#1}\else
  \providecommand{\doi}{doi:\discretionary{}{}{}\begingroup \urlstyle{rm}\Url}\fi

\bibitem[{Agarwal and Manuel(2011)}]{agarwal2011incorporating}
Agarwal, P. and Manuel, L. (2011).
\newblock Incorporating irregular nonlinear waves in coupled simulation and reliability studies of offshore wind turbines.
\newblock \emph{Applied Ocean Research}, 33(3), 215--227.

\bibitem[{Carmo et~al.(2024)Carmo, Bergua, Wang, and Robertson}]{carmo2024validation}
Carmo, L., Bergua, R., Wang, L., and Robertson, A. (2024).
\newblock Validation of local structural loads computed by openfast against measurements from the focal experimental campaign.
\newblock In \emph{International Conference on Offshore Mechanics and Arctic Engineering}, volume 87851, V007T09A048. American Society of Mechanical Engineers.

\bibitem[{Duarte et~al.(2014)Duarte, Sarmento, and Jonkman}]{duarte2014effects}
Duarte, T.M., Sarmento, A.J., and Jonkman, J.M. (2014).
\newblock Effects of second-order hydrodynamic forces on floating offshore wind turbines.
\newblock In \emph{32nd ASME Wind Energy Symposium}, 0361.

\bibitem[{Faltinsen(1993)}]{faltinsen1993sea}
Faltinsen, O. (1993).
\newblock \emph{Sea loads on ships and offshore structures}, volume~1.
\newblock Cambridge university press.

\bibitem[{Hasan et~al.(2023)Hasan, Sarker, Ngo, and Das}]{hasan2023stabilization}
Hasan, T., Sarker, D., Ngo, T., and Das, T. (2023).
\newblock Stabilization of the wind turbine floating platform using mooring actuation.
\newblock \emph{IFAC-PapersOnLine}, 56(3), 535--540.

\bibitem[{H{\o}eg and Zhang(2023)}]{hoeg2023semi}
H{\o}eg, C.E. and Zhang, Z. (2023).
\newblock A semi-analytical hydrodynamic model for floating offshore wind turbines (fowt) with application to a fowt heave plate tuned mass damper.
\newblock \emph{Ocean Engineering}, 287, 115756.

\bibitem[{Ishihara and Zhang(2019)}]{ishihara2019prediction}
Ishihara, T. and Zhang, S. (2019).
\newblock Prediction of dynamic response of semi-submersible floating offshore wind turbine using augmented morison's equation with frequency dependent hydrodynamic coefficients.
\newblock \emph{Renewable energy}, 131, 1186--1207.

\bibitem[{MacCamy and Fuchs(1954)}]{maccamy1954wave}
MacCamy, R.C. and Fuchs, R.A. (1954).
\newblock \emph{Wave forces on piles: a diffraction theory}.
\newblock 69. US Beach Erosion Board.

\bibitem[{Noboni et~al.(2025)Noboni, McConnell, and Das}]{noboni2025modeling}
Noboni, T., McConnell, J., and Das, T. (2025).
\newblock Modeling tethered multirotor autogyro with altitude control via differential rotor braking.
\newblock \emph{Journal of Guidance, Control, and Dynamics}, 1--14.

\bibitem[{Roald et~al.(2013)Roald, Jonkman, Robertson, and Chokani}]{roald2013effect}
Roald, L., Jonkman, J., Robertson, A., and Chokani, N. (2013).
\newblock The effect of second-order hydrodynamics on floating offshore wind turbines.
\newblock \emph{Energy Procedia}, 35, 253--264.

\bibitem[{Robertson(2023)}]{robertson2023focal}
Robertson, A. (2023).
\newblock Focal campaign iv: Integrated system control: Turbine+ hull.
\newblock Technical report, Pacific Northwest National Lab.(PNNL), Richland, WA (United States~….

\bibitem[{Robertson et~al.(2014)Robertson, Jonkman, Masciola, Song, Goupee, Coulling, and Luan}]{robertson2014definition}
Robertson, A., Jonkman, J., Masciola, M., Song, H., Goupee, A., Coulling, A., and Luan, C. (2014).
\newblock Definition of the semisubmersible floating system for phase ii of oc4.
\newblock Technical report, National Renewable Energy Lab.(NREL), Golden, CO (United States).

\bibitem[{Sarker et~al.(2024{\natexlab{a}})Sarker, Hasan, Ngo, and Das}]{sarker2024causality}
Sarker, D., Hasan, T., Ngo, T., and Das, T. (2024{\natexlab{a}}).
\newblock Causality-free modeling and validation of a semisubmersible floating offshore wind turbine platform with tuned mass dampers.
\newblock \emph{IEEE Journal of Oceanic Engineering}, 49(4), 1430--1454.

\bibitem[{Sarker et~al.(2024{\natexlab{b}})Sarker, Tran, Mohsin, Odeh, Ngo, and Das}]{sarker2024modeling}
Sarker, D., Tran, D., Mohsin, K., Odeh, M., Ngo, T., and Das, T. (2024{\natexlab{b}}).
\newblock Modeling, validation, and control of the iea-15mw reference wind turbine and volturnus-s platform.
\newblock \emph{IFAC-PapersOnLine}, 58(28), 1--6.

\bibitem[{Wang et~al.(2023{\natexlab{a}})Wang, Bergua, Robertson, Jonkman, Ngo, Das, Sarker, Flavia, Harries, Fowler et~al.}]{wang2023experimental}
Wang, L., Bergua, R., Robertson, A., Jonkman, J., Ngo, T., Das, T., Sarker, D., Flavia, F.F., Harries, R., Fowler, M., et~al. (2023{\natexlab{a}}).
\newblock Experimental validation of models of a hull-based tuned mass damper system for a semisubmersible floating offshore wind turbine platform.
\newblock \emph{Journal of Physics: Conference Series}, 2626(1), 012067.

\bibitem[{Wang et~al.(2022{\natexlab{a}})Wang, Jonkman, Hayman, Platt, Jonkman, and Robertson}]{wang2022recent}
Wang, L., Jonkman, J., Hayman, G., Platt, A., Jonkman, B., and Robertson, A. (2022{\natexlab{a}}).
\newblock Recent hydrodynamic modeling enhancements in openfast.
\newblock In \emph{International Conference on Offshore Mechanics and Arctic Engineering}, volume 86618, V001T01A004. American Society of Mechanical Engineers.

\bibitem[{Wang et~al.(2025)Wang, Robertson, Jonkman, Liao, Berthelsen, Abdelmoteleb, Rohrer, Rajasree, Bachynski-Poli{\'c}, Clement et~al.}]{wang2025oc7}
Wang, L., Robertson, A., Jonkman, J., Liao, Y., Berthelsen, P.A., Abdelmoteleb, S.E., Rohrer, P., Rajasree, V.R.N., Bachynski-Poli{\'c}, E., Clement, C., et~al. (2025).
\newblock Oc7 phase i: Toward practical sea-state-dependent modeling of hydrodynamic viscous drag and damping.
\newblock \emph{Ocean Engineering}, 336, 121745.

\bibitem[{Wang et~al.(2022{\natexlab{b}})Wang, Robertson, Jonkman, and Yu}]{wang2022oc6}
Wang, L., Robertson, A., Jonkman, J., and Yu, Y.H. (2022{\natexlab{b}}).
\newblock Oc6 phase i: Improvements to the openfast predictions of nonlinear, low-frequency responses of a floating offshore wind turbine platform.
\newblock \emph{Renewable Energy}, 187, 282--301.

\bibitem[{Wang et~al.(2023{\natexlab{b}})Wang, Moan, and Gao}]{wang2023methodology}
Wang, S., Moan, T., and Gao, Z. (2023{\natexlab{b}}).
\newblock Methodology for global structural load effect analysis of the semi-submersible hull of floating wind turbines under still water, wind, and wave loads.
\newblock \emph{Marine Structures}, 91.

\bibitem[{Zhang et~al.(2020)Zhang, Shi, Karimirad, Michailides, and Jiang}]{zhang2020second}
Zhang, L., Shi, W., Karimirad, M., Michailides, C., and Jiang, Z. (2020).
\newblock Second-order hydrodynamic effects on the response of three semisubmersible floating offshore wind turbines.
\newblock \emph{Ocean Engineering}, 207, 107371.

\end{thebibliography}
\end{document}